\documentclass[aps,twocolumn,groupedaddress,showkeys,showpacs]{revtex4}
\usepackage[dvips]{graphicx}
\begin{document}

\title{Analytic study of the three-urn model for separation of sand}

\author{G.\ M.\ Shim}
\author{B.\ Y.\ Park}
\author{J.\ D.\ Noh}
\author{Hoyun Lee}
\affiliation{Department of Physics, Chungnam National University, 
Daejeon 305-764, Korea}

\date{\today}

\begin{abstract}
We present an analytic study of the three-urn model for separation of sand.
We solve analytically the master equation and the first-passage problem. 
We find that the stationary probability distribution obeys the detailed balance 
and is governed by the {\it free energy}.
We find that the characteristic lifetime of a cluster diverges algebraically 
with exponent $1/3$ at the limit of stability.
\end{abstract}


\maketitle

\section{INTRODUCTION}

A granular system consisting of macroscopic particles exhibits extremely 
rich phenomena, which has been studied both experimentally and 
theoretically~\cite{Kadanoff}. One of such interesting phenomena is the 
spatial separation of shaken sand.
In the experiment by Schlichting and Nordmeier~\cite{Schlichting96}, 
granular particles are prepared in a box which is mounted on a shaker and 
separated into two equal parts by a wall. There is a slit on a wall through 
which particles can move from one compartment to the other.
Under a certain shaking condition, the granular particles simultaneously 
separate into dense and dilute regions, which will not occur for gaseous 
particles, due to the dissipative nature of the macroscopic particle 
collision.

The emergence of symmetry breaking in shaken sand was first explained by
Eggers using a hydrodynamic approach~\cite{Eggers99}, and later by 
Lipowski and Droz using an urn model which is supposed to 
capture the essence of the experimental system~\cite{Lipowski02a}. 
In the urn model, $N$ granular particles are distributed into $L$
urns. And each particle can jump from one urn to another with the
probability controlled by a parameter, called an effective temperature,
which depends explicitly on the density of particles in each urn.
The dissipative nature of the particle collision is incorporated into the
model with the density-dependent effective temperature.

The urn model is simple enough to allow extensive numerical 
calculations~\cite{Lipowski02a,Coppex02}
as well as analytical studies~\cite{Shim03,Bena03}.
The two-urn~($L=2$) model shows a rich structure with
symmetric, mixed, and asymmetric phases separated with continuous 
and discontinuous transitions as well as the tricritical point.
In the symmetric phase particles are distributed equally in each urn,
while the symmetry is broken in the asymmetric phase. In the mixed phase,
both the symmetric and asymmetric states are stable.
It was also found that the characteristic time which it takes to reach 
the symmetric state from an asymmetric state is given by 
the same free energy function which governs the stationary probability
distribution~\cite{Shim03}.
Coppex et al.\ also numerically investigated the three-urn model to
find the absence of continuous transition~\cite{Coppex02}. 
They also obtained that the characteristic time at the limit of stability
diverges algebraically as the number of particles increases~(see also
Ref.~\cite{vanderMeer}).

In this paper, we present the results of an analytic study to the 
master equation and the characteristic time scale in the three-urn model.
We solve the master equation in the thermodynamic limit to find that the
solution shows a nature of deformed wave.
We also obtain the stationary probability distribution in a finite system,
which interestingly obeys the detailed balance. Combining the knowledge of
the characteristic scales of those distribution and the mean-field flux
equation, we obtain the scaling property of the characteristic time scale
at the limit of stability.

The paper is organized as follows. 
In Sec.~\ref{sec:II} we briefly review the model and its master equation. 
In Sec.~\ref{sec:III} we present the analytic solution of the master 
equation in the thermodynamic limit and the analytic expression of 
the stationary probability distribution.
In Sec.~\ref{sec:IV} we investigate the scaling law of the
characteristic time scale.
Section \ref{sec:V} is devoted to conclusions and discussions.

\section{MODEL AND ITS MASTER EQUATION} \label{sec:II}

The model introduced by Coppex et al.\ \cite{Coppex02} is 
defined as follows. $N$ particles are distributed between three urns, 
and the number of particles in each urn is denoted as $N_1$, $N_2$, 
and $N_3=N-N_1-N_2$, respectively. 
At each time of updates one of the $N$ particles is randomly chosen. 
Let $n$ be a fraction of the total number of particles in the urn 
which the selected particle belongs to. 
With probability $\exp(-\frac{1}{T(n)})$ the selected particle moves 
to a randomly chosen neighboring urn. $T(n) = T_0+\Delta(1-n)$ is the 
effective temperature of an urn with particles $nN$ that measures the 
strength of fluctuations in the urn. 
For more detailed description of the model,
we refer readers to Ref.~\cite{Lipowski02a}.

It is easy to derive the master equation for the probability
distribution $p(N_1,N_2,N_3,t)$ that there are $N_i$ particles in 
urn $i$ at time $t$
\begin{eqnarray} 
  && p(N_1,N_2,N_3,t+1) = 
  \nonumber \\
  &&\hspace{1em} F\bigl(\frac{N_1+1}{N}\bigr) \frac12\biggl(
                 p(N_1+1,N_2-1,N_3,t)
         \nonumber \\
         &&\hspace{5em} +p(N_1+1,N_2,N_3-1,t) \biggr)
  \nonumber\\
  &&\hspace{1em}  + F\bigl(\frac{N_2+1}{N}\bigr) \frac12\biggl(
                         p(N_1-1,N_2+1,N_3,t)
         \nonumber \\
         &&\hspace{5em} +p(N_1,N_2+1,N_3-1,t) \biggr)
  \nonumber\\
  &&\hspace{1em}  + F\bigl(\frac{N_3+1}{N}\bigr) \frac12\biggl(
                         p(N_1-1,N_2,N_3+1,t)
         \nonumber \\
         &&\hspace{5em} +p(N_1,N_2-1,N_3+1,t) \biggr)
  \nonumber\\
  &&\hspace{1em}  + \Bigl[  1-\sum_{i=1}^3 F\bigl(\frac{N_i}{N}\bigr)
                 \Bigr]\ p(N_1,N_2,N_3,t)
\,,\label{eq:master}\end{eqnarray}
where $F(n) = n \exp(-\frac{1}{T(n)})$ measures the flux of 
particles leaving the given urn.
Here we introduced for convenience the notation $p(N_1,N_2,N_3,t)=0$ 
if $N_1+N_2+N_3 \ne N$ or any $N_i$'s is either less than 0 or greater 
than $N$.

Let's denote the occupancies of the urns by $n_i = N_i/N$.
The time evolution of the averaged particle occupancies
is governed by the equations
\begin{equation}\label{eq:averagedoccupancy}
   \langle n_i\rangle_{t+1} = \langle n_i\rangle_t 
   +\frac1{N} \langle {\cal F}_i(\vec{n}) \rangle_t
\,,\end{equation}
where $\langle \cdots \rangle_t = 
\sum_{N_1,N_2,N_3} (\cdots) p(N_1,N_2,N_3,t)$, 
and ${\cal F}_i(\vec{n}) = \frac12\sum_{k=1}^3F(n_k)-\frac32F(n_i)$.
The unit of time may be chosen so that there is a single update per 
a particle on average. Scaling the time by $N$, expanding 
Eq.~(\ref{eq:averagedoccupancy}) with respect to $\frac{1}{N}$, 
and using the mean-field approximation in evaluating the average, we get
\begin{equation}\label{eq:differentialoccupancy}
  \frac{d}{dt}n_i(t) =  {\cal F}_i(\vec{n}(t))
\,.\end{equation}
where $n_i(t) = \langle n_i \rangle_t$. 

\begin{figure}[tbp]
   \includegraphics*[width=\columnwidth]{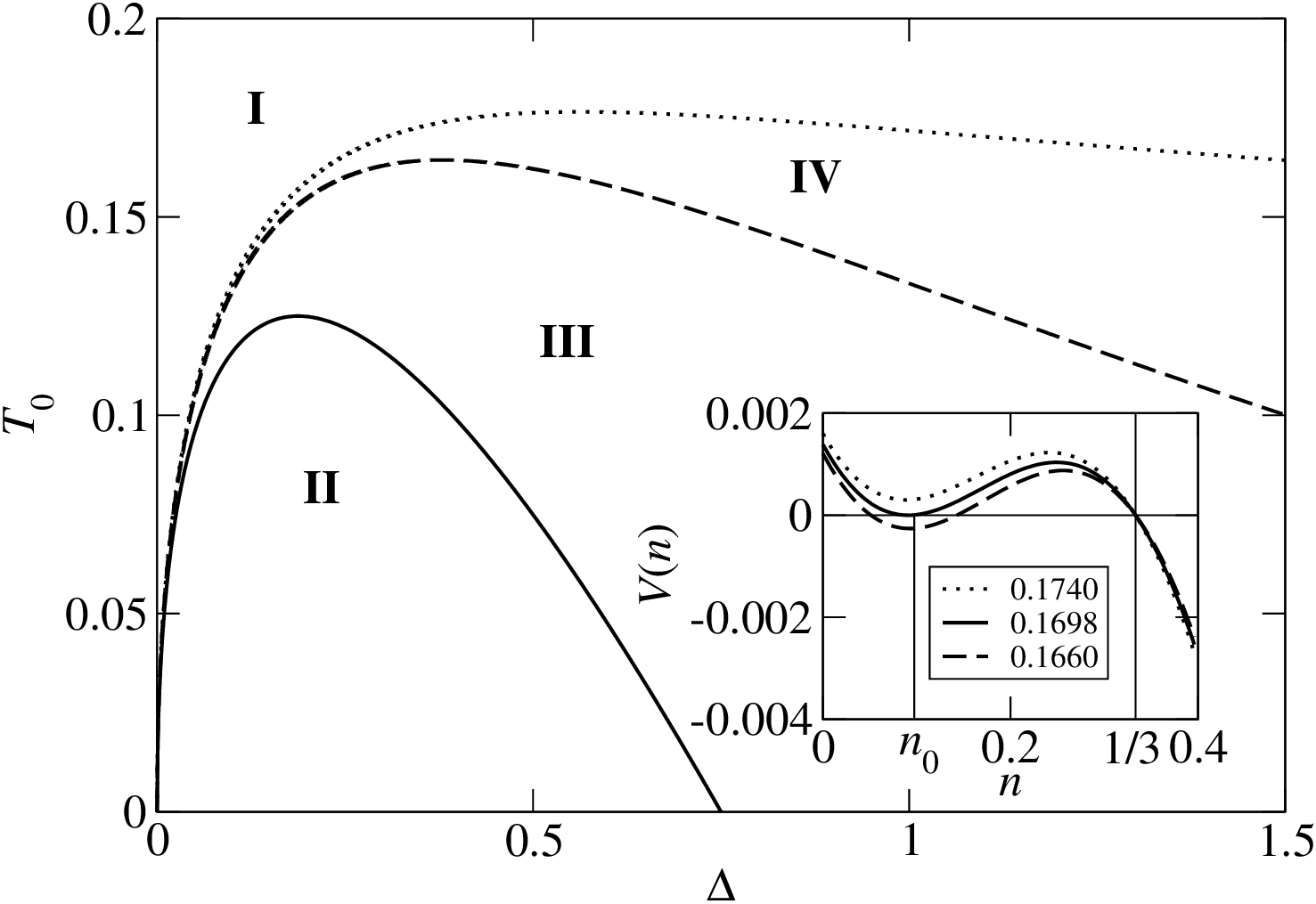} 
   \caption{\label{fig:phasediagram} Phase diagram of the three-urn model
            \cite{Coppex02}.
            The symmetric solution vanishes continuously on the solid line
            while the asymmetric one disappears discontinuously on the 
            dotted line. The transition of the behavior of the stationary
            probability distribution is denoted by the dashed line. For
            the inset, see Sec.~\ref{sec:IV}.}
\end{figure}

Detailed analysis on the existence of the stable stationary solutions of
Eq.~(\ref{eq:differentialoccupancy}) was done by Coppex et al. \cite{Coppex02}.
We here display their phase diagram in Fig.~\ref{fig:phasediagram} to make 
our paper as self-contained as possible. 
The stable symmetric solution ($n_1=n_2=n_3$) exists in region I, III, and
IV while the stable asymmetric solution ($n_1>n_2=n_3$) exists in region II,
III and IV. Note that
Eq.~(\ref{eq:differentialoccupancy}) is obtained with mean-field
approximation that $\langle {\cal F}(\vec{n})\rangle = {\cal F}(\langle
\vec{n}\rangle)$. However, as will be shown later, the relation becomes
exact for the delta-peaked initial probability distribution, so is the phase diagram.

\section{THE SOLUTION OF THE MASTER EQUATION} \label{sec:III}

We are mainly interested in investigating the properties of the 
infinite system.
Consider the thermodynamic limit $N \rightarrow \infty$ with
the occupancies of the urns $n_i$ being fixed. 
Representing the probability distribution by $n_i$ instead of $N_i$, 
scaling the time by $N$, 
expanding Eq.~(\ref{eq:master}) with respect to $\frac{1}{N}$, 
and keeping the terms up to the first order, 
we arrive at the partial differential equation
\begin{equation}\label{eq:continuousmaster}
    \frac{\partial}{\partial t}p(\vec{n},t)
   + \frac{\partial}{\partial n_1}\bigl[ 
              {\cal F}_1(\vec{n})p(\vec{n},t) \bigr]
   + \frac{\partial}{\partial n_2}\bigl[ 
              {\cal F}_2(\vec{n})p(\vec{n},t) \bigr]
   = 0
\,.\end{equation}
Here we would like to remind that $n_3 = 1-n_1-n_2$ so that the independent
variables are $n_1, n_2$.

\begin{figure}[tbp]
   \includegraphics*[width=7cm]{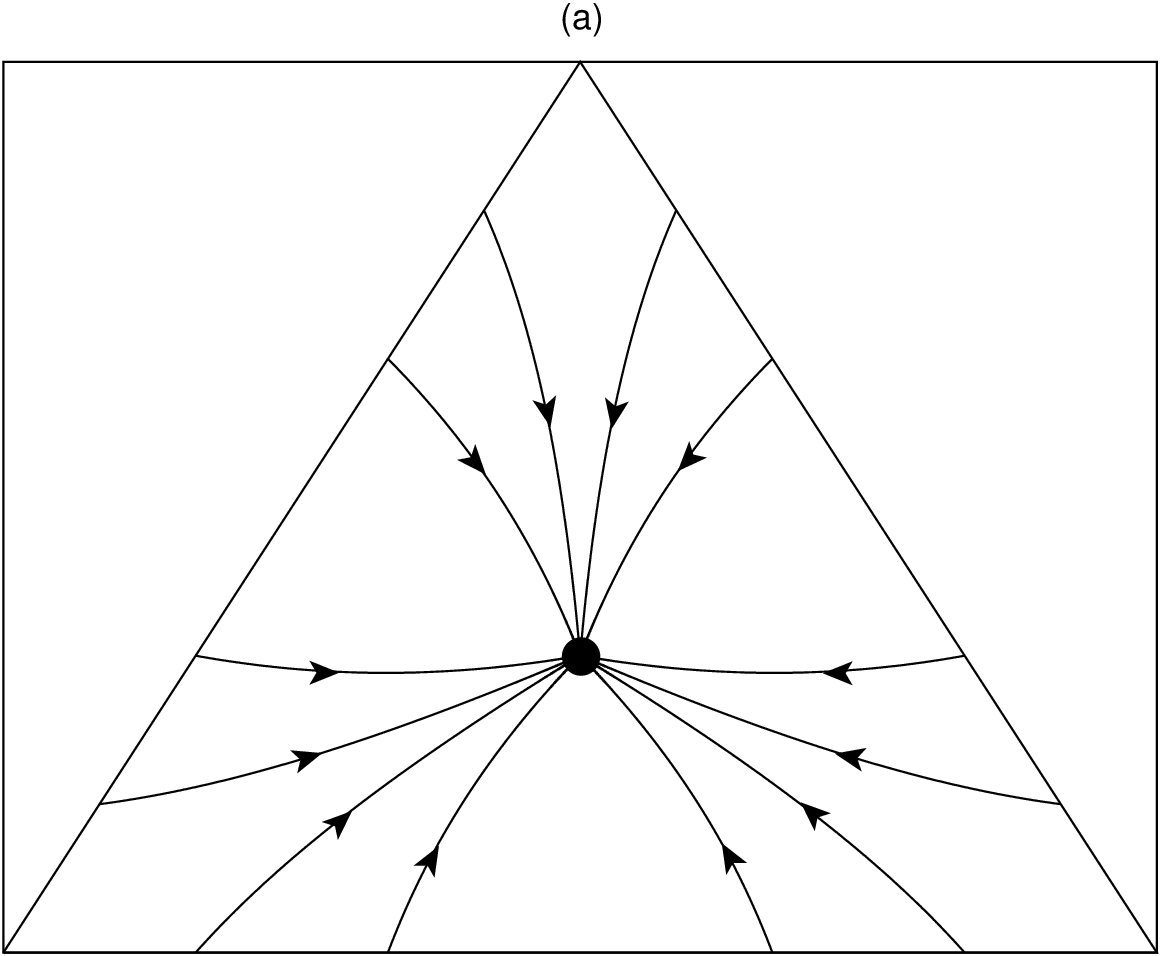} 
   \includegraphics*[width=7cm]{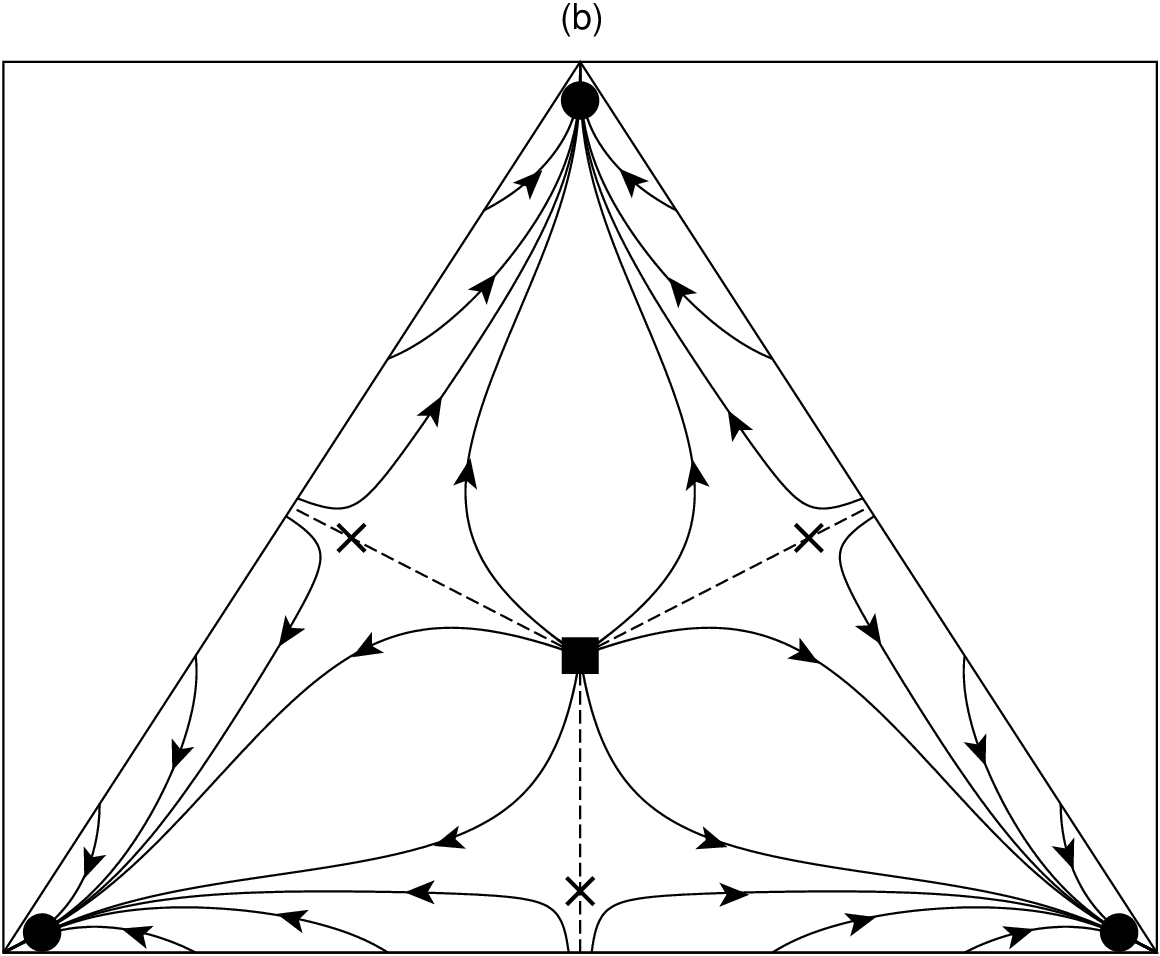} 
   \includegraphics*[width=7cm]{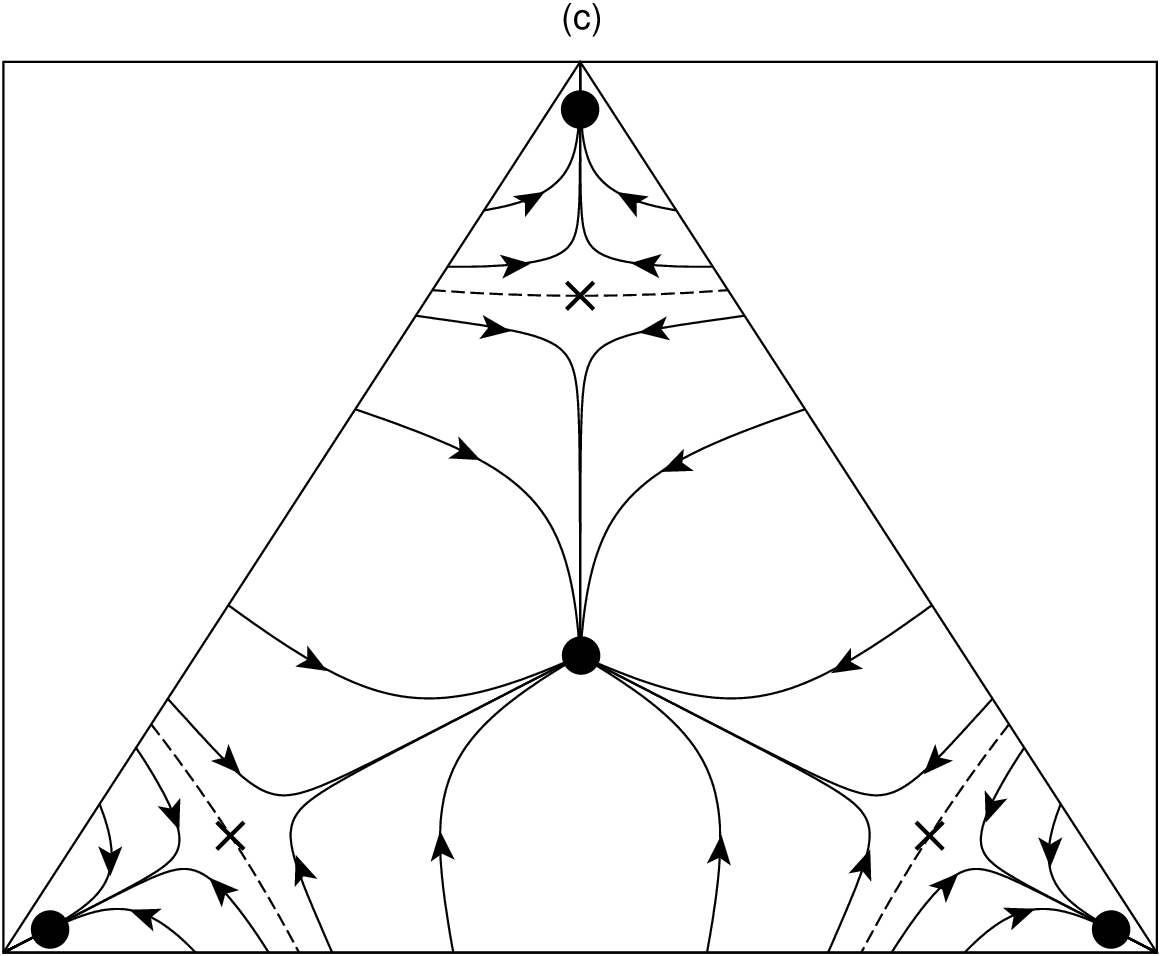} 
   \caption{\label{fig:flow} Typical trajectories of
    Eq.~(\ref{eq:differentialoccupancy}). 
   We plotted the trajectories in the diagonal plane of a unit cube 
   so that each variable $n_i$ are treated in the same way. 
   (a) $\Delta=0.5, T_0=0.5$ (symmetric phase) 
   (b) $\Delta=0.25, T_0=0.05$ (asymmetric phase) 
   (c) $\Delta=1.0, T_0=0.1$ (mixed phase).
   The stable, the unstable, and the saddle point are represented with a filled
   circle, a filled square, and a cross, respectively.}
\end{figure}

Note that Eq.~(\ref{eq:continuousmaster}) may be interpreted as the continuity 
equation for the probability with velocity $({\cal F}_1,{\cal F}_2)$.
Since the velocity is already given as a function of $\vec{n}$, the
solution to Eq.~(\ref{eq:continuousmaster}) can be formally found as
follows. Time evolution of a point located at $\vec{r}=(x,y,1-x-y)$ at $t=0$ 
is determined by the Eq.~(\ref{eq:differentialoccupancy}). We denote it by 
$\vec{R}(t;\vec{r})$.
We show its typical trajectories in Fig.~\ref{fig:flow}.
In the symmetric phase (region I in Fig.~\ref{fig:phasediagram}) there is 
only the stable fixed point corresponding to the symmetric state so that 
every trajectory flows to that point. This is shown in Fig.~\ref{fig:flow}(a).
Figure \ref{fig:flow}(b) shows a typical behavior for the asymmetric
phase (region II in Fig.~\ref{fig:phasediagram}). In this case, there are
three stable fixed points corresponding to asymmetric states and one
unstable fixed points corresponding to the symmetric state as well as
three saddle points. The trajectories are separated by the separatrices 
denoted by dashed lines.
Finally Fig.~\ref{fig:flow}(c) shows a
typical behavior for the mixed phase (region III and IV in
Fig.~\ref{fig:phasediagram}). There are one stable symmetric fixed point
and three asymmetric fixed points as well as three saddle points.
The trajectories flowing toward the stable fixed points are separated
by the separatrices.

Since the continuity equation implies that the probability at point 
$\vec{r}$ evolves according to $\vec{R}(t;\vec{r})$,
we obtain a formal solution of Eq.~(\ref{eq:continuousmaster})
\begin{equation}\label{eq:formalsolution}
 p(\vec{n},t) = \int dxdy\, p(\vec{r},0)
                \delta^{(2)}\biggl(\vec{n}-\vec{R}(t;\vec{r})\biggr)
\,.\end{equation}
It implies that the initial probability distribution in the basin of
attraction associated with a stable fixed point will eventually 
accumulate at that point. As a consequence, in the long time limit
$t\rightarrow\infty$ the probability distribution becomes a sum of delta
peaks at the stable fixed points denoted by $\vec{n}_i$;
\begin{equation}\label{eq:probdistinfty}
    p(\vec{n},\infty) = \sum_i p_i \delta^{(2)}(\vec{n}-\vec{n}_i)
\,,\end{equation}
where $p_i$ are the sum of the initial probabilities in the basin of 
attraction associated with $\vec{n}_i$.

Now we will consider another limit in the master equation 
(\ref{eq:master}), namely take the long time limit $t \rightarrow
\infty$ before we take the limit $N \rightarrow \infty$. 
That is, we are looking for the stationary probability distribution for a
finite system.  Though this limit may not properly reflect the properties 
of the infinite system, we expect it will reveal interesting properties
of the system concerning the characteristic times (See \cite{Shim03} in
two-urn case.). 

Let's first take the long time limit of $t \rightarrow \infty$ in 
Eq.~(\ref{eq:master}). In this limit we may drop off the time dependence in
the probability distribution, which now takes the form
\begin{eqnarray} 
\left[ F(\frac{N_1+1}{N}) \frac{p(N_1+1,N_2-1,N_3)}{p(N_1,N_2,N_3)} 
        - F(\frac{N_2}{N}) \right] &+& \nonumber \\
\left[ F(\frac{N_2+1}{N}) \frac{p(N_1,N_2+1,N_3-1)}{p(N_1,N_2,N_3)} 
        - F(\frac{N_3}{N}) \right] &+& \nonumber \\
\left[ F(\frac{N_3+1}{N}) \frac{p(N_1-1,N_2,N_3+1)}{p(N_1,N_2,N_3)} 
        - F(\frac{N_1}{N}) \right] &+& \nonumber \\
\left[ F(\frac{N_1+1}{N}) \frac{p(N_1+1,N_2,N_3-1)}{p(N_1,N_2,N_3)} 
        - F(\frac{N_3}{N}) \right] &+& \nonumber \\
\left[ F(\frac{N_3+1}{N}) \frac{p(N_1,N_2-1,N_3+1)}{p(N_1,N_2,N_3)} 
        - F(\frac{N_2}{N}) \right] &+& \nonumber \\
   \left[ F(\frac{N_2+1}{N}) \frac{p(N_1-1,N_2+1,N_3)}{p(N_1,N_2,N_3)} 
        - F(\frac{N_1}{N}) \right] &=& 0 
\!.\label{eq:stationaryprob}\end{eqnarray}

In contrast to the corresponding equation (13) in the two-urn model
\cite{Shim03}, it does not show a simple tridiagonal structure. 
However, we can show that the stationary probability distribution
determined by Eq.~(\ref{eq:stationaryprob}) obeys the {\it detailed balance}
\begin{equation}\label{eq:detailedbalance}
   F\bigl(\frac{N_i+1}{N}\bigr)p(N_1',N_2',N_3') 
   =
   F\bigl(\frac{N_j}{N}\bigr)p(N_1,N_2,N_3) 
\,,\end{equation}
where $\{i,j,k\}=\{1,2,3\}$ and $N_i'=N_i+1, N_j'=N_j-1,N_k'=N_k$.
We also would like to note that this kind of relation is obeyed in
two-urn model.
Equation (\ref{eq:stationaryprob}) essentially tells us that probability
distribution with $N_j$ is given by that with $N_j-1$, which is in turn
given by that with $N_j-2$, and so on. Repeatedly using
Eq.~(\ref{eq:stationaryprob}) with $i$ as 1 or 2 and $i=3$, we obtain 
\begin{equation}\label{eq:stationarysolution}
  p(N_1,N_2,N_3) = \frac{
                   \prod_{k=1}^{N_1+N_2}F\bigl(\frac{N-k+1}{N}\bigr)
                  }{ 
                   \prod_{i=1}^{N_1}F\bigl(\frac{i}{N}\bigr)
                   \prod_{j=1}^{N_2}F\bigl(\frac{j}{N}\bigr)
                  }p(0,0,N)
\,,\end{equation}
where $p(0,0,N)$ appears as an overall factor to normalize the
probabilities so that we get
$$
       p(0,0,N) = \biggl[1+\sum_{N_1=0}^N\sum_{N_2=0}^{N-N_1} \frac{
                   \prod_{k=1}^{N_1+N_2}F\bigl(\frac{N-k+1}{N}\bigr)
                  }{ 
                   \prod_{i=1}^{N_1}F\bigl(\frac{i}{N}\bigr)
                   \prod_{j=1}^{N_2}F\bigl(\frac{j}{N}\bigr) }
                  \biggr]^{-1} \,.
$$

Now let's take the limit $N \rightarrow \infty$.
With $\frac{N_1}{N} = n_1$, $\frac{N_2}{N}=n_2$, 
and scaling the probability distribution by
$N^2$, the stationary probability distribution for large $N$ now becomes
\begin{equation}\label{eq:statcontinprob}
   p(\vec{n}) \approx \frac{ e^{NG(\vec{n})} }{
                 \int_0^1 dx \int_0^{1-x} dy \, e^{NG(\vec{r})} }
\,,\end{equation}
where 
\begin{eqnarray}
    G(\vec{n}) &=& \int_0^{n_1+n_2} dt \, \bigl[ \ln F(1-t) \bigr]
  \nonumber \\ \label{eq:nfe}
               && - \int_0^{n_1} dt \, \bigl[ \ln F(t) \bigr]
                  - \int_0^{n_2} dt \, \bigl[ \ln F(t) \bigr]
\,.\end{eqnarray}
We will call $G(\vec{n})$ the (negative) free energy function.

In the limit $N \rightarrow \infty$, the main contribution to the
stationary probability distribution comes only from the maximum of
$G(\cdot)$, and it becomes the sum of delta peaks. 
The maximum of $G(\vec{n})$ occurs when 
\begin{equation}
   \frac{\partial}{\partial n_1}G(\vec{n})
  =\frac{\partial}{\partial n_2}G(\vec{n})
  =0
\end{equation}
or
\begin{equation}
      F(n_1) = F(n_2) = F(n_3)
\,.\end{equation}
This condition is equivalent to the stationary condition of the flux
equation (\ref{eq:differentialoccupancy}).  Note that in region II only 
the three stable asymmetric solutions having the same maximum  
exist, while in region I only the symmetric solution is stable. 
Therefore the stationary probability distribution has the triple peaks 
in region II and  only the central peak in region I. 
In region III and IV both the
symmetric and the asymmetric solutions are stable so that the maximum
of $G(\vec{n})$ should be determined by comparing its values at the
stable fixed points. The crossover of the maximum point occurs when
both values coincide. This implies that the transition between
the triple peaks and the central single peak in the probability
distribution is determined by the condition
$\Delta G = G(\vec{n}_a)-G(1/3,1/3,1/3) = 0$, where $\vec{n}_a$ is one of 
the stable asymmetric fixed points. 
This condition yields a line that separates two regions III and IV.

\section{CHARACTERISTIC TIME Scale}\label{sec:IV}
One of the interesting phenomena in sands separated by compartments is 
a sudden collapse of a granular cluster~\cite{vanderMeer,Coppex02}.
It is observed experimentally that a granular cluster with majority 
of sand grains in a single compartment remains stable for a long time
until it collapses abruptly and diffuses over all
compartments~\cite{vanderMeer}. The urn model proposed in
Ref.~\cite{Coppex02} displays the same phenomenon: In region I 
a granular cluster is stable up to a time scale $\tau$, after which sand
particles are distributed uniformly over all boxes. Approaching the phase
boundary I-IV, the characteristic time scale diverges. At the phase
boundary, it is found numerically that $\tau$
scales as $\tau \sim N^z$ with $z\simeq 0.32$~\cite{Coppex02}.
In this section, the scaling law for the characteristic time $\tau$
at and near the phase boundary is derived analytically.

The master equation in Eq.~(\ref{eq:continuousmaster}) in the large $N$
limit does not contain a diffusive term. It implies that a delta-peaked 
probability distribution remains delta-peaked during time evolution, which
enabled us to write down the formal solution in 
Eq.~(\ref{eq:formalsolution}). The dispersion-less nature also guarantees
that the mean field approximation in Eq.~(\ref{eq:differentialoccupancy})
for the occupancy $n_i$ becomes exact as long as the initial values of 
$n_i$'s are prescribed. Hence we can study the dynamics of the granular
cluster using Eq.~(\ref{eq:differentialoccupancy}) with the initial
condition $n_1=n_2=0$ and $n_3=1$.

With $n_1=n_2=n$ and $n_3 = 1-2n$, one can rewrite 
Eq.~(\ref{eq:differentialoccupancy}) as
\begin{equation}
\dot{n} = V(n) \equiv \frac{1}{2}\left\{F(1-2n)-F(n)\right\}  
\end{equation}
with the initial condition $n(t=0)=0$.
The cluster dynamics is then determined from the property of the flow
function $V(n)$. In the inset of Fig.~\ref{fig:phasediagram}, 
we show the plots of the
flow function at different values of $T_0$ with $\Delta=0.3$ fixed. At
$\Delta=0.3$, the critical point separating the region I and VI is given by 
$T_0 = T_{0c} = 0.1698\cdots$. For $T_0>T_{0c}$, $n=1/3$ is the unique 
stationary state solution; $n$ grows from zero to 1/3 to reach the symmetric
state with $n_1=n_2=n_3=1/3$. Note that there exists a local minimum in
$V(n)$ at $0<n_0<1/3$ where the flow velocity is minimum.
Hence the value of $n$ remains at the intermediate value $n\simeq n_0$
for a long time, and then converges to $n=1/3$ quickly. 
It turns out that the sudden collapse of a granular
cluster~\cite{Coppex02,vanderMeer} is due to the local minimum in $V(n)$.
The characteristic time scale or the life time $\tau$ is given by
\begin{equation}\label{eq:lifetime}
\tau = \int_0^{n_0} \frac{dn}{V(n)} \ .
\end{equation}
As $T_0\rightarrow T_{0c}^+$, the minimum flow velocity $V(n_0)$ decreases
and hence $\tau$ increases. 
The asymmetric solution with $n_1=n_2<n_3$ emerges at $T=T_{0c}$ when 
$V(n_{0})=0$ and $\tau$ diverges.

The scaling law for the characteristic time scale is determined from the
analytic property of $V(n)$ near $T_0\simeq T_{0c}$ and $n\simeq n_0$.
One can expand as $V(n) \simeq  a (n-n_0)^2 + b (T_0 - T_{0c})$,
where $a$ and $b$ are constants of order one and all higher order terms in
$(T_0-T_{0c})$ are irrelevant. Inserting it into Eq.~(\ref{eq:lifetime}),
one obtains that the characteristic time scale diverges as $\tau\sim
(T_0-T_{0c})^{-1/2}$~(see also Ref.~\cite{vanderMeer}). 
At $T_0 = T_{0c}$, the granular system approaches the asymmetric state
algebraically in time as $|n-n_0| \sim t^{-1}$ whose life time $\tau$
diverges. That is to say, the asymmetric state is stable in the 
$N\rightarrow \infty$ limit. 

However, for finite $N$, the system could evolve into the symmetric state 
at $T_0 = T_{0c}$ with the help of statistical fluctuations in $n$.
Indeed, the probability distribution $p$ has a finite dispersion for finite
$N$, even though it becomes a sum of delta peaks for infinite $N$.
In order to estimate the order of magnitude of fluctuations in $n$, we
investigate the stationary probability distribution in 
Eq.~(\ref{eq:statcontinprob}) at $T_0 = T_{0c}$ in detail. Near the
asymmetric state with $n_1=n_2=n$ and $n_3 = 1-2n$, the (negative) free 
energy function in Eq.~(\ref{eq:nfe}) is written as $G(n) = \int_0^{2n}dt
\ln F(1-t) - 2 \int_0^n dt \ln F(t)$. Now we expand it near $n=n_0$ to yield
$G(n) = G(n_0) + G'(n_0)(n-n_0) + \frac{1}{2}G''(n_0) (n-n_0)^2 +
\frac{1}{3} G'''(n_0) (n-n_0)^3 + \cdots$. After straightforward
calculations, one can show that 
\begin{eqnarray}
G'(n_0) &=& 2\ln (F(1-2n_0)/F(n_0)) \\
G''(n_0) &=& \frac{2}{F(n_0)}V'(n_0) \\
G'''(n_0) &=& \frac{2}{F(n_0)}V''(n_0) \ .
\end{eqnarray} 
Note that $G'(n_0)=G''(n_0)=0$ since $V(n_0) =
V'(n_0)=0$. Therefore, the (negative) free energy function can be
approximated as $G(n) \simeq c (n-n_0)^3$ with a constant $c$, and hence the
probability distribution as $p_s (n) \simeq e^{cN(n-n_0)^3}$ up to a
normalization constant. 

The stationary probability distribution suggests that the magnitude of
the fluctuation in $n$ near $n_0$ is of order $\delta n \sim N^{-1/3}$.
With the help of the fluctuation, the granular system could get away from
the asymmetric state when $|n(t) - n_0| \sim \delta n$, and flow into the 
symmetric state. Therefore the characteristic time scale is given by 
$\tau \simeq \int_0^{n_0-\delta n} dt / V(t)$, which results in $\tau \sim N^z$
with the dynamic exponent
\begin{equation}\label{eq:dynamicexponent}
z = 1/3 \ .
\end{equation}
Our analytic result confirms the numerical result $z \simeq 0.32$ reported
in Ref.~\cite{Coppex02}.


\section{CONCLUSIONS}\label{sec:V}

We analytically investigate the three-urn model introduced by Coppex
et al.\ \cite{Coppex02}
We formally solve the master equation of the
model in the thermodynamic limit and find how the probability
distribution evolves. 
In the long time limit, the probability distribution becomes 
delta peaks only at the stable fixed points. 
The strength of a delta peak is equal to the sum of initial 
probabilities in the basin of attraction associated with that fixed
point.

We solve exactly the stationary probability distribution where we
take the long time limit before we take thermodynamic limit.
We find the distribution obeys the detailed balance.
Regardless of the initial probability distribution it shows triple
peaks or a single central peak depending on the parameters of the
system. The final formula of the stationary probability distribution 
resembles that of the equilibrium systems, where the transition from 
the triple peaks to the single peak is determined by the condition 
that {\it free energies} of two phases become equal.

We also obtain the exact scaling law for the characteristic time scale
$\tau$ which it takes to reach the symmetric state from an asymmetric 
state near and at the phase boundary I-VI. In the symmetric phase~(region 
I), the granular cluster is unstable and  $\tau$ is finite. It grows as one
approach the phase boundary as $\tau \sim (T_0 - T_{0c})^{-1/2}$.
At $T_0 = T_{0c}$, the characteristic time diverges algebraically as $\tau
\sim N^z$ with the dynamic exponent $z=1/3$.


\end{document}